# Coordinated Initial Access in Millimetre Wave Standalone Networks

Yinan Qi and Maziar Nekovee
Samsung Electronics R&D Institute UK
Communications House, South Street, Staines, Middlesex TW18 4QE, UK

*Abstract*— In this paper, a novel coordinated initial access (IA) scheme for clustered millimeter wave small cells (mmSCs) is proposed for the fifth generation mobile communication networks (5G). This solution is a method for efficient and fast initial access for ultra-dense millimeter wave standalone networks without presence of overlaid legacy networks operating on lower frequency. In contrast to the current full beam sweep scheme, where time consuming exhaustive searching is employed, the mmSCs within one cluster will perform the IA procedure in a coordinated manner based on the power delay profile (PDP) measurement reports shared with each other via the backhaul links and thereby avoiding the full beam sweep. The proposed scheme significantly reduces the initial access time, enhances the access robustness and reduces the cost and complexity of the mobile terminals.

*Keywords—millimeter wave, 5G, beamforming, initial access (IA)*

## I. INTRODUCTION

There has been a significant growth in the volume of mobile data traffic in recent years due to proliferation of smart phones and other mobile devices that support a wide range of broadband applications and services. In order to support user data rates of Gbps and above, mobile communication systems operating in higher frequencies, e.g., millimeter wave (mm-wave) bands, draw more and more attention from both industry and academia as a very promising approach due to the potential of utilizing the much larger spectrum bandwidth available in mm-wave frequency bands [1]-[3].

However, due to the hostile propagation condition in mm-wave radio channels, e.g., severe path loss, vulnerability to blockage, etc., large antenna gains at both transmitter and receiver sides are required to overcome propagation losses [4]-[6]. In this regard, very large scale antenna arrays are needed that enable highly directive transmit and receive beamforming. As a result of highly directive transmission and reception in mm-wave communication, the cell discovery becomes more challenging than omni-directional transmission and reception. In the conventional cellular networks, such as 3GPP LTE/LTE-A, the antenna pattern of the base station (BS) is sectorized and the antenna pattern of the user equipment (UE) is normally assumed to be omni-directional [7]. When the UE tries to access the radio access network (RAN), it sends out a preamble during the random access (RA) procedure. The reception of the RA preamble is mainly affected by the distance between the BS and the UE. However, the high directivity in mm-wave bands makes the conventional initial access procedures unsuitable. Therefore, the initial access procedure needs to be designed properly to exploit the resources in the spatial dimension. More specifically, to detect/discover a BS in the proximity of a UE, both the mm-wave transmitter and receiver need to align their transmission and reception beams with each other [8].

There are two important key performance indicators (KPIs) to evaluate the efficiency of initial access: 1) the discovery/initial access time between the initiation of cell discovery and its completion and 2) the beam sweep signaling overhead. Without any a priori knowledge about the UE, e.g., location, the IA procedure is not a trivial task but an exhaustive combinatorial problem. In the exhaustive approach, all beam pairs are examined by sending a training packet for each beam pair [9]. As a result of this, the discovery time could become prohibitively long.

The IA time can be reduced in a non-standalone scenario [10], where mm-wave network inter-networks with another overlaid legacy network to explore the context information of the UE to accelerate the IA procedure with mmSCs. The focus of this paper is also on the design of the IA procedure with reduced IA time. However, the fundamental assumption of network architecture is different by aiming at standalone mm-wave networks where assistance from overlaid networks is not available. The cost efficiency of the UE can be improved since dual RF interfaces are no longer needed and the signaling overhead exchanged between the mm-wave network and the overlaid legacy network is also avoided. The rest of the paper is organized as follows. Some related works will be introduced in the next section and the new coordinated IA approach is proposed and analyzed in section III. In section IV, numerical results are presented to illustrate the performance improvement and the final section concludes the paper.

## II. RELATED WORKS

Fig.1 illustrates the IA procedure between a single UE and a single mmSC in existing mm-wave technologies [8]-[9]. We assume pre-defined beam codebooks for simplicity. As can be seen, the UE needs to conduct full beam sweep for a given mmSC receive (Rx) beam and each possible Rx beam should be examined. For example, if there are $N_{rx}$ beams at the mmSC and $N_{tx}$ beams at the UE, the IA time could add up to $N_{tx}N_{rx}T_{RA}$, where $T_{RA}$ is the examining time for one beam pair.

This exhaustive beam sweep method not only causes a very long IA time, but it also introduces significant signaling overhead since the preamble needs to be transmitted repeatedly. Moreover, the repeated transmission increases the energy consumption of the UE thereby draining its battery.

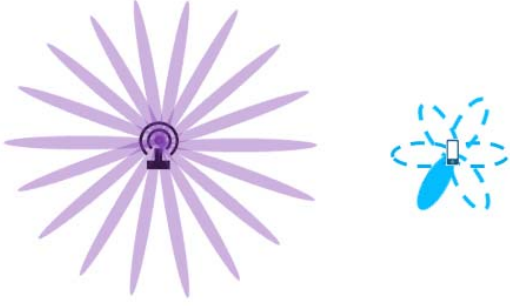

Fig. 1 Initial access of a single UE to a single mmSC

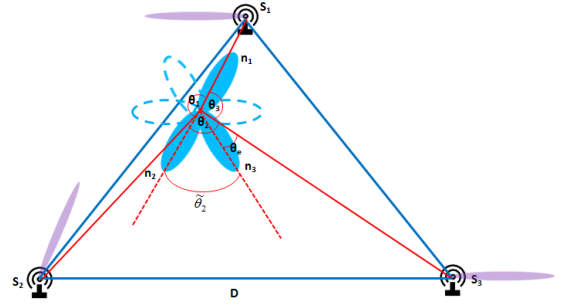

Fig. 2 System model

The initial access procedure can be facilitated by exploring the concept of user/control plane splitting in a non-standalone network, where the mm-wave network inter-networks with another overlaid legacy network operating in a lower frequency [10]. The UE is assumed to have dual RF interfaces for legacy network link and mm-wave link, respectively, and the UE has already established a UL connection with a legacy macro cell. The control plane is separated from user plan and located in the overlaid legacy network where the control signaling and context information of a UE can be conveyed. The user plane is located in the mmSC network, i.e., once the UE requires high rate data transmission, it needs to establish connections with one or multiple mmSCs. In order to do so, the UE measures the DL channel from an mmSC using the reference signals and generates a measurement report. When the IA procedure is initiated, the UE sends the measurement report to the legacy macro cell via the established links. Based on the measurement reports, the macro cell sends a recommended Rx beam set to the mmSC via backhaul links. The mmSCs then reorder the Rx beams to put the recommended beams in the forepart. The IA time is reduced at the cost of significantly increased signaling overhead.

### III. Coordinated Initial Access Design

Our work effectively builds on the clustered mmSCs scenario, where the mmSCs within a cluster are connected with each other via the backhaul links. In this paper we assume that the mmSCs are allowed to coordinate by sharing the power delay profile (PDP) measurement reports via backhaul links in addition to the location and beam codebook information. Based on the measurement reports, the IA procedure can be coordinated jointly in a more sophisticated manner to effectively reduce the IA time.

#### A. System Model

The proposed solution involves at least three mmSCs and can be divided into four phases: measurement, coordinated beam sweep reordering, initial access, and asymmetric multi-cell association. In the first phase the involved mmSCs measure the PDP of the received preamble and share the measurement reports via the backhaul links. In the beam sweep reordering phase, mmSCs jointly estimate the best Rx beam for each mmSC based on the measurement reports and reorder the mmSC Rx beams to put the best Rx beams in forepart. In the third phase, the UE conduct beam sweep for given Rx beam triplet until an uplink connection is established successfully. In the last phase, the existing uplink connection can be used to establish multiple uplink connections which may or may not associate the UE to the same mmSC in the downlink.

For the purpose of simplicity of presentation, we assume that there are three mmSCs in one cluster forming an equilateral triangle with side length $D$. The extension to other triangles is straightforward. The UE is assumed to be located in the triangle. Again, it can be easily extended to the case where the UE is located outside the triangle. As shown in Fig. 2, the distances between the UE and the mmSCs are assumed to be $d_1$, $d_2$ and $d_3$ and angles are assumed to be $\theta_1$, $\theta_2$, and $\theta_3$.

Since the mmSCs are connected via the backhaul links, they can be synchronized to start Rx beam sweep at the same time. The Rx beam directions are randomly chosen at the beginning when the UE conduct the first full beam sweep.

We can compute the signal received by the $i$th mmSC as

$$P_{sc} = P_{UE} + G_{UE}(\varphi_{UE,i}) + G_{sc}(\varphi_{sc,i}) - L(d_i) \quad (1)$$

where $P_{UE}$ is the transmit power of the UE, $G_{UE}$ and $G_{sc}$ are the UE and mmSC antenna gains, respectively, $\varphi_{UE,i}$ is the angle between the main UE antenna lobe and the $i$th mmSC, $\varphi_{sc,i}$ is the angle between the main antenna lobe of the $i$th mmSC and the UE, $L$ is the pathloss component, and $d_i$ is the distance between the $i$th mmSC and the UE.

A simplified directional antenna model is assumed for both the mmSC and the UE [11], given as

$$G(\varphi) = \begin{cases} G_0 - 3.01 \times \left( \dfrac{2\varphi}{\varphi_{-3dB}} \right)^2, & 0 \leq \varphi \leq \varphi_{ml}/2 \\ G_{sl}, & \varphi_{ml}/2 \leq \varphi \leq \pi \end{cases} \quad (2)$$

where

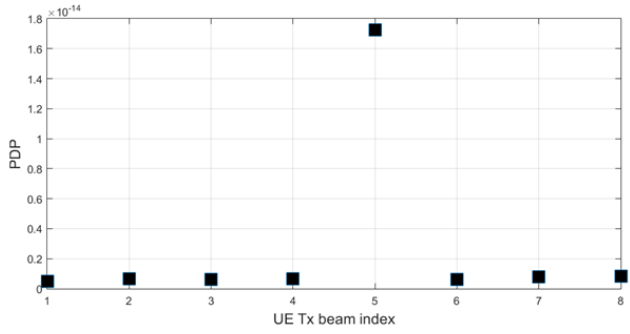

Fig. 3   PDP values for different UE beam index

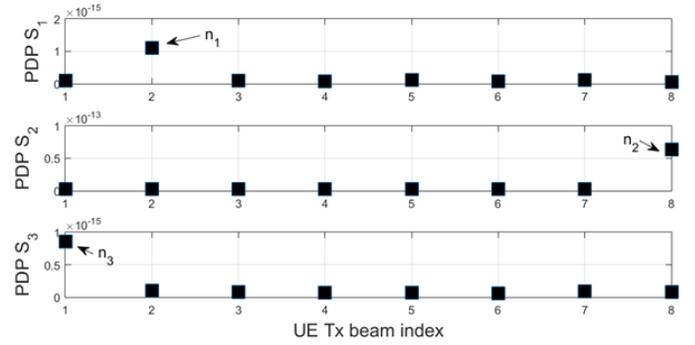

Fig. 4   PDP of 3 mmSCs

$$\phi_{ml} = 2.6\phi_{-3dB},$$

$$G_0 = 10\log\left(\frac{1.6162}{\sin(\phi_{-3dB}/2)}\right)^2, G_{sl} = -0.4111\ln(\phi_{-3dB}) - 10.579.$$

Here $\varphi$ is an arbitrary angle within the range $[0, \pi]$, $\varphi_{-3dB}$ is the angle of half-power beamwidth, $\varphi_{ml}$ is the main lobe width in units of degrees, and $G_0$ and $G_{sl}$ are the maximum antenna gain and the side lobe gain, respectively.

The pathloss model is given by [12]

$$L(d) = 61.4 + 21\log(d) \quad (3)$$

where $d$ is the distance in meters.

### B. PDP Measurement

In LTE, UE broadcasts a preamble sequence built by cyclicly-shifting a Zadoff-Chu (ZC) sequence of prime length $N_{zc}$ [13],

$$x_u(n) = \exp\left[-j\frac{\pi u n(n+1)}{N_{ZC}}\right], 0 \leq n \leq N_{ZC} - 1 \quad (4)$$

where $u$ is the index of the sequence. The preamble sequence will be transmitted using SC-FDMA and the detailed IA and detection procedures can be found in [7]. We assume that the same ZC sequences and IA procedure are employed and the PDP of the received sequence is

$$\mathrm{PDP}(l) = |z_u(l)|^2 = \left|\sum_{n=0}^{N_{zc}-1} y(n)x_u^*(n+l)\right|^2 \quad (5)$$

where $y(n)$ is the received preamble sequence and $z_u(l)$ is the discrete periodic correlation function at lag $l$. Once PDP peaks above a detection threshold $\gamma_{ra}$, which is defined based on the target false detection probability are found, we simply assume a mmSC is discovered. There might be some possibility of collision and failure in the procedures to fully establish the connection, but these are out of the scope of this paper.

As aforementioned, due to the hostile propagation environment the beam directions of the mmSC and the UE need to be aligned so that the PDP peaks can be detected. For a given beam direction of the mmSC, which may or may not point to the UE, if UE conducts full sweep, i.e., transmits a preamble in every possible direction, a measurement report can be generated, where for each UE beam index a PDP peak is obtained. The PDP peaks for different UE beam indices are illustrated in Fig. 3. As can be seen, the PDP peak of UE beam index $i$ (here $i = 5$) is significantly larger than others because the direction of this transmitting (Tx) beam points to the mmSC. This measurement report will be shared within the mmSC cluster. For a cluster with at least 3 mmSCs, the UE's location can be coarsely estimated based on the measurement reports (details in the next sub-section).

### C. Best Beam Estimation

The mm-wave SCs beam directions are chosen randomly initially and the UE conducts full beam sweep and transmits the preamble for each Tx beam index every $T_{ra}$ seconds. Each mmSC calculates PDP values per $T_{ra}$ seconds and after $N_{tx}T_{ra}$ seconds one Tx full sweep is completed and a measurement report is generated at each mmSC. Within each measurement report, there are $N_{tx}$ PDP peaks, each corresponding to one UE Tx beam index. One mmSC can acquire three measurement reports exchanged via the backhaul links as shown in Fig. 4. Three peak values of the measurement reports identify three UE Tx beam index $n_1$, $n_2$ and $n_3$. The difference of the Tx beam index $n_1$, $n_2$ and $n_3$ can be used to approximate $\theta_i$ in Fig. 2 as

$$\tilde{\theta}_i = \begin{cases} \dfrac{2\pi(n_{i+1} - n_i)}{N_{tx}}, & \text{if } n_{i+1} > n_i \\ \dfrac{2\pi(N_{tx} - n_i + n_{i+1})}{N_{tx}}, & \text{if } n_{i+1} < n_i \end{cases} \quad (6)$$

It should be noted that (6) is only an approximation because the central direction of the UE Tx beam may not be perfectly aligned with the direction of the mmSCs as shown in Fig. 2. The estimation accuracy can be improved with narrower UE beam width. However, since the number of antennas installed at the UE is limited, we should not assume very narrow UE beam width. With approximated $\theta_i$, the distance between the UE and the mmSCs can be easily estimated by solving the following equations:

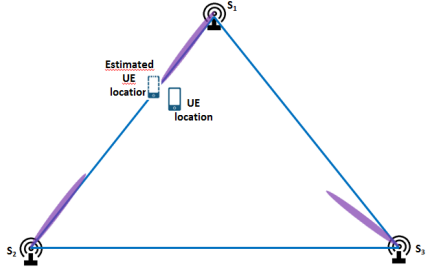

Fig. 5  mmSC Rx beams reordering based on estimated UE location

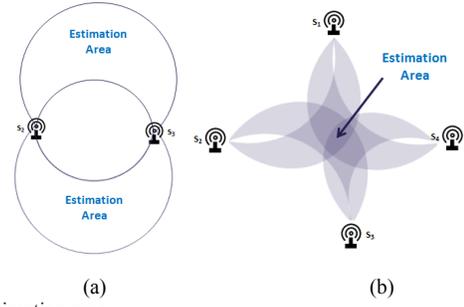

Fig. 6  Estimation areas

$$\begin{cases} d_1^2 + d_2^2 - 2d_1 d_2 \cos(\hat{\theta}_1) = D^2 \\ d_2^2 + d_3^2 - 2d_2 d_3 \cos(\hat{\theta}_2) = D^2 \\ d_1^2 + d_3^2 - 2d_1 d_3 \cos(\hat{\theta}_3) = D^2 \end{cases} \quad (7)$$

With $d_1$, $d_2$ and $d_3$, the location of the UE can be approximated. The mmSCs can therefore reorder the Rx beam set and choose the Rx beams pointing at the estimated UE location in the following UE Tx beam sweep as shown in Fig. 5. As can be seen, there is possibility that not all three chosen Rx beams point to the actual UE location because of the estimation error, but there is a high probability that at least one or two Rx beams point to the correct direction. Therefore, in the second UE Tx beam sweep, there is a larger chance to pair the Rx and Tx beams.

If the cluster size is larger than 3, we can simply choose three measurement reports with the largest PDP peaks and the rest of the estimation procedure remains the same. By doing this, we actually further improve the robustness of the IA approach. We can also use more than three measurement reports to further enhance the estimation accuracy. Considering the sharp drop of the antenna gain when $\varphi > \varphi_{ml}$, we can assume that the PDP peaks can only be achieved when $\varphi \leq \varphi_{ml}$. Then instead of a single-valued estimation of $\theta_i$, we can estimate the range of $\theta_i$ as

$$\tilde{\theta}_i - \varphi_{ml} \leq \theta_i \leq \tilde{\theta}_i + \varphi_{ml} \quad (8)$$

If the angle between one point to two fixed points is constant, the loci of this point are circular arcs that pass through two fixed points [14]. Therefore the following equations identify an area, denoted as estimation area as shown in Fig. 6 (a)

$$\begin{aligned} d_1^2 + d_2^2 - 2d_1 d_2 \cos(\theta_1) &= D^2 \\ \tilde{\theta}_i - \varphi_{ml} &\leq \theta_i \leq \tilde{\theta}_i + \varphi_{ml} \end{aligned} \quad (9)$$

With three mmSCs, the UE's position is supposed to be in the overlapping area of three estimation areas identified by three corresponding equations in (7). With additional mmSCs, we have an additional equation and thus the additional estimation areas. The overlapping of all estimation areas can be further refined as shown in Fig. 6 (b).

### D. NLOS Consideration

The proposed best beam estimation only applies to the scenario that there are LOS links between the mmSCs and the UE. However, the LOS link between the UE and the mmSC could be blocked so that the UE beam actually points to a reflector and thereby forming a NLOS link. In such a case, the estimated UE position could be different from its actual location and therefore none of the three mmSCs can appropriately adjust their Rx beams towards the UE. However, the NLOS links can be efficiently avoided if the mmSC cluster consists of more than 3 mmSCs. Assuming the reflection coefficient is 0.7, i.e., the NLOS link is at least 1.5dB worse than the LOS link [15], Monte Carlo simulation is conducted. The blocking probability of each mmSC is denoted as $P_{blk}$ and three mmSCs with the largest PDP peak are chosen from the cluster to estimate the UE position.

The probability that the three chosen mmSCs are not blocked, i.e., they have LOS links to the UE, is depicted in Fig. 7. As can be seen, for a small block probability $P_{blk}=0.1$, the probability that LOS links exist, i.e., $P_{LOS}$, is more than 0.9 when the cluster size is larger than 10. Even with a large block probability $P_{blk}=0.5$, $P_{LOS}$ can be above 0.7 with the cluster size larger than 20. In a highly densified deployment scenario, cluster size could be even larger than 20 and thus there is a large chance that the LOS links exist.

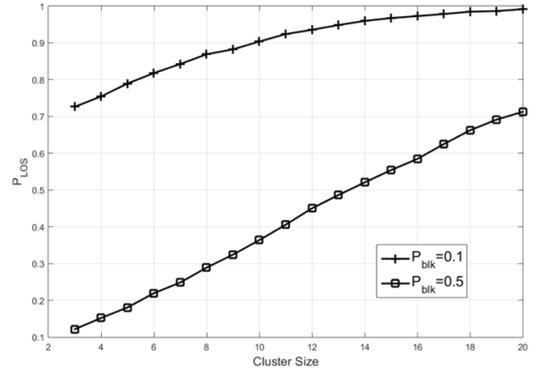

Fig. 7  $P_{LOS}$ = probability that LOS links exist

### E. Other Issues

As aforementioned, the robustness of the proposed initial access approach depends on the level of deployment densification. The requirement of three mmSCs conducting preamble reception simultaneously might be a luxury situation in certain cases but the deployment of small cells is expected to be highly densified in 5G to ensure that the proposed approach could work properly.

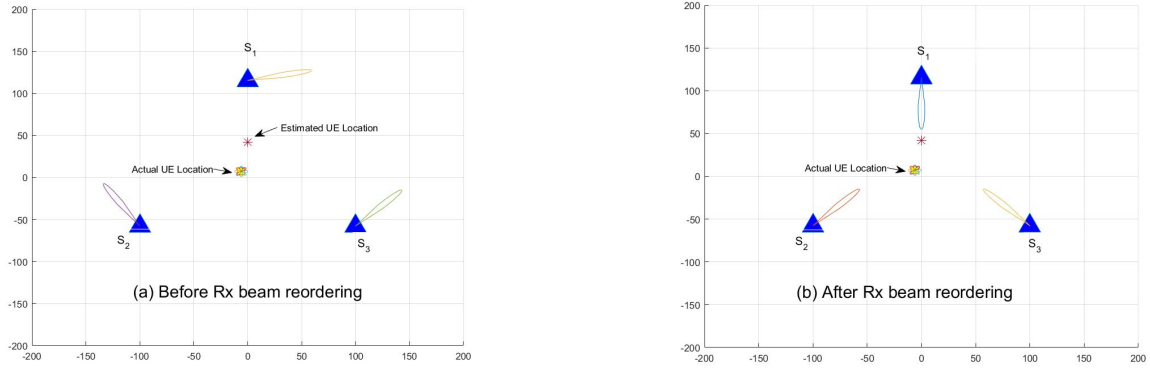

Fig. 8  UE location estimation

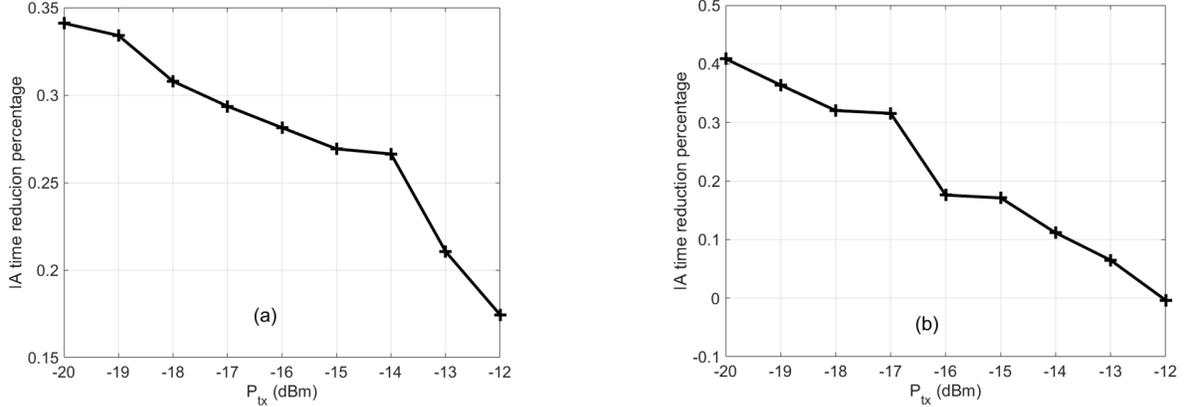

Fig. 9  IA time reduction percentage (a)$N_{tx}$=4 (b)$N_{tx}$=8

In addition, the measurement report only needs a few bits to carry the index of the beam with PDP peak and thus only causes negligible signaling overhead in backhaul network.

## IV. SIMULATION RESULTS

In this section, we present simulation results obtained based on the proposed IA approach. The detailed system parameters are presented in Table I. We compare our approach with a benchmark approach, where each mmSC does independent Rx beam sweep without any coordination. The initial access procedure is claimed to be successful once the UE is connected to any mmSC.

Fig. 8 (a) shows the estimated UE location using the proposed approach. As can be seen, after one round of UE Tx full beam sweep, its location can be estimated and the estimated location is quite close to the actual location. In the next beam sweep round, the Rx beam from $S_1$ will point at the UE as shown in Fig. 8 (b).

Fig. 9 shows the percentage of IA access time reduction for UE Tx beam number $N_{tx}$=4 and 8, respectively, given by:

Table-I System Parameters

| Parameters | Value |
|---|---|
| Carrier Frequency | 28GHz |
| Bandwidth | 1.08MHz |
| $N_0$ | -171 dBm/Hz |
| Inter-mmSC distance | 200 m |
| Length of ZC sequence | 839 |

$$P_{er} = \frac{T_{ra\_new} - T_{ra\_con}}{T_{ra\_con}} \times 100\%, \quad (10)$$

where $T_{ra\_new}$ and $T_{ra\_con}$ are IA time of the proposed approach and the conventional one, respectively. As can be seen, IA time decreases with increased UE Tx power, which is reasonable since with very high UE Tx power, the PDP of received sequence can be so strong that it exceeds the threshold even without Tx and Rx beam paring. It can also be seen that IA time is significantly reduced with proposed invention, especially when the transmission power of the UE is not high, which is normally the case in mm-wave systems. Fig. 10 shows the percentage of IA time reduction against miss detection probability $P_{miss}$ for $N_{tx}$=4 and 8, respectively. With a larger target $P_{miss}$, IA time can be further reduced. With 1% target $P_{miss}$, which is normally used in LTE/LTE-A, IA time can be reduced by 22% and 18% for $N_{tx}$=4 and 8, respectively, using the proposed approach. Fig. 11 shows the normalized IA time when $N_{sc}$, i.e., the size of the cluster changes. For $N_{sc}$=1, conventional exhausting searching is employed. For $N_{sc} \geq 3$, the proposed approach is employed and additional mmSCs are utilized to further refine the approximation. It can be seen that there is a significant reduction once the proposed approach is employed. Moreover, using additional SCs can further reduce the IA time.

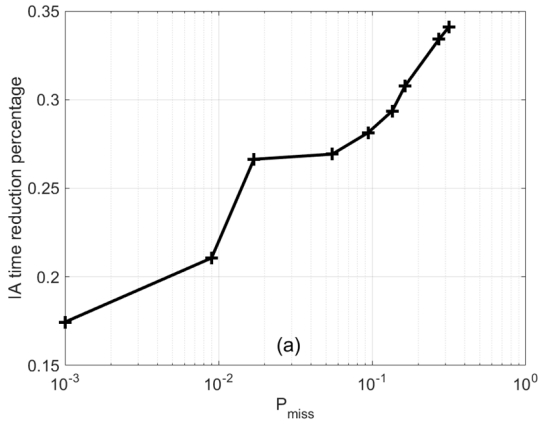 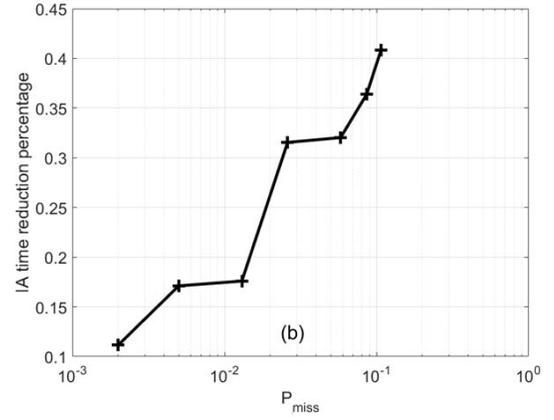

Fig. 10  IA time reduction percentage (a)$N_{tx}$=4 (b)$N_{tx}$=8

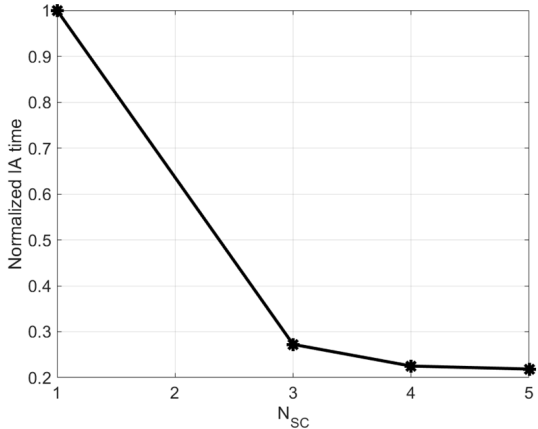

Fig. 11  Normalized IA time with increased cluster size

## V. CONCLUSIONS

In this paper, a novel coordinated IA scheme for standalone mm-wave networks is proposed. In contrast to discovery mechanisms relying on the overlaid legacy networks, the proposed approach aims at standalone mm-wave networks and does not require assistance from overlaid networks, which is a step forward towards standalone operation. The signaling overhead between the mm-wave network and overlaid legacy network is also avoided. The proposed scheme significantly reduces the beam discovery time of the IA procedure, enhances the robustness, and reduces the cost and complexity of the mobile terminals.